\documentclass[12pt]{article}
\usepackage{epsfig}
\newcommand{\be}{\begin{equation}}
\newcommand{\ee}{\end{equation}}
\newcommand{\bdm}{\begin{displaymath}}
\newcommand{\edm}{\end{displaymath}}
\renewcommand{\thefootnote}{\fnsymbol{footnote}}
\def\simlt{\mathrel{\lower2.5pt\vbox{\lineskip=0pt\baselineskip=0pt
           \hbox{$<$}\hbox{$\sim$}}}}
\def\simgt{\mathrel{\lower2.5pt\vbox{\lineskip=0pt\baselineskip=0pt
           \hbox{$>$}\hbox{$\sim$}}}}
\newcommand{\ls}[1]
   {\dimen0=\fontdimen6\the\font
    \lineskip=#1\dimen0
    \advance\lineskip.5\fontdimen5\the\font
    \advance\lineskip-\dimen0
    \lineskiplimit=.9\lineskip
    \baselineskip=\lineskip
    \advance\baselineskip\dimen0
    \normallineskip\lineskip
    \normallineskiplimit\lineskiplimit
    \normalbaselineskip\baselineskip
    \ignorespaces}

\catcode`@=11
\newcount\@tempcntc
\def\@citex[#1]#2{\if@filesw\immediate\write\@auxout{\string\citation{#2}}\fi
  \@tempcnta\z@\@tempcntb\m@ne\def\@citea{}\@cite{\@for\@citeb:=#2\do
    {\@ifundefined
       {b@\@citeb}{\@citeo\@tempcntb\m@ne\@citea\def\@citea{,}{\bf ?}\@warning
       {Citation `\@citeb' on page \thepage \space undefined}}%
    {\setbox\z@\hbox{\global\@tempcntc0\csname b@\@citeb\endcsname\relax}%
     \ifnum\@tempcntc=\z@ \@citeo\@tempcntb\m@ne
       \@citea\def\@citea{,}\hbox{\csname b@\@citeb\endcsname}%
     \else
      \advance\@tempcntb\@ne
      \ifnum\@tempcntb=\@tempcntc
      \else\advance\@tempcntb\m@ne\@citeo
      \@tempcnta\@tempcntc\@tempcntb\@tempcntc\fi\fi}}\@citeo}{#1}}
\def\@citeo{\ifnum\@tempcnta>\@tempcntb\else\@citea\def\@citea{,}%
  \ifnum\@tempcnta=\@tempcntb\the\@tempcnta\else
   {\advance\@tempcnta\@ne\ifnum\@tempcnta=\@tempcntb \else \def\@citea{--}\fi
    \advance\@tempcnta\m@ne\the\@tempcnta\@citea\the\@tempcntb}\fi\fi}
\catcode`@=12

\newcommand{\zp}{Z. Phys.\ {\bf C}}

\begin{document}
\setcounter{footnote}{1}
\begin{flushright}
\today
\end{flushright}
\vspace{7mm}
\begin{center}
\Large{{\bf 
Flavor Violation in  Warped Extra Dimensions \\
and CP Asymmetries in 
$B$ Decays 
}}
\end{center}
\vspace{5mm}
\begin{center}
{\large\bf Gustavo Burdman\\
\vspace{0.3cm}
{\normalsize\it 
Theoretical Physics  Group, 
Lawrence Berkeley National Laboratory  
Berkeley, CA 94720}
}
\end{center}
\vspace{0.50cm}
\thispagestyle{empty}
\begin{abstract}
We show that  CP asymmetries 
in $b\to s$  hadronic decays are potentially affected by 
the presence of massive color-octet particles strongly coupled
to the third generation quarks. 
Theories with warped extra dimensions provide
natural candidates in the Kaluza-Klein excitations of gluons in 
scenarios where
flavor-breaking by bulk fermion masses 
results in the localization of  fermion wave-functions. 
Topcolor models, in which a new gauge interaction leads to top-condensation 
and a large top mass, also result in the presence of these color-octet states
with TeV masses. 
We find that large effects are possible
in modes such as $B\to\phi K_s$, $B\to \eta' K_s$ and 
$B\to\pi^0 K_s$ among others. 
\end{abstract}
\newpage

\renewcommand{\thefootnote}{\arabic{footnote}}
\setcounter{footnote}{0}
\setcounter{page}{1}

\section{Introduction}
It is generally believed that extensions of the 
standard model must address the stability of the weak scale 
with new physics at energies not far above the TeV scale. 
The origin of fermion masses, on the other hand, could in principle reside
at much higher scales. Nonetheless, it is tempting to consider the possibility
that at least part of the fermion masses is dynamically generated at similar
scales. This is particularly true when considering the origin of the top
quark mass, which is of the order of the weak scale. The correspondingly
large Yukawa coupling suggests the presence of a strongly coupled sector
associated with the third generation quarks. 

Although originally proposed as a solution to the hierarchy problem~\cite{rs1},  
the existence of warped extra dimensions provides a class of scenarios
that may also address fermion masses as dynamical in origin. 
The Anti de-Sitter metric
in a compact extra dimension is responsible for generating a low
energy scale (the weak scale) on one of the fixed points, even when there is 
only one fundamental scale ($\simeq M_P$). Although in the earlier versions of this
scenario matter and gauge fields were localized on the infrared (TeV) brane, 
it was soon realized the potential for model building of 
introducing gauge and matter fields in the
bulk~\cite{bulk1,chang,gn}. In particular, it was recognized in 
Refs.~\cite{gp,huber1} that in non-supersymmetric scenarios, the 
localization of fermion zero modes, which is controlled by order one parameters
in the bulk theory, could be used to explain the fermion mass hierarchy. 
Zero-mode fermions localized toward the TeV brane would have large
overlaps with the TeV-localized Higgs field, and therefore an order one
Yukawa coupling. On the other hand, the Yukawa couplings of fermions localized 
on the Planck brane would be exponentially suppressed. Order one flavor breaking
in the bulk fermion masses results in an exponentially enhanced hierarchy in the
zero-mode Yukawa couplings. On the other hand, and as it is pointed out in 
Ref.~\cite{gp}, the couplings of the zero-mode fermions to the lightest
Kaluza-Klein (KK) excitations of gauge bosons become large for TeV brane localized 
fermions. 
Although this might result in tight constraints from electroweak precision 
measurements~\cite{dhr1,dhr2}, the actual bounds are quite model dependent, with 
the most important factors being the choice of ``weak'' gauge group in the 
bulk and the localization of the light fermions. However, even if the latter are considered
on the Planck brane in order to avoid electroweak bounds, at least part of the 
third generation should be TeV localized.
This constitutes a potentially large source for flavor violation, 
particularly with third generation quarks. 
       
Such dynamics is also present in purely four dimensional theories. 
For instance, this is the case in Topcolor theories~\cite{topin}, 
where a gauge  interaction strongly coupled to the third generation quarks 
is spontaneously broken at the TeV scale. 
Topcolor theories have to either be supplemented
by some additional dynamics (as in Ref.~\cite{tcatc}) or 
by additional matter as in the top see-saw models~\cite{topss}. 
However, a rather model independent
statement that can be made about these theories, is that they will result in 
non-universal interactions with a color-octet massive gauge boson, which 
in turn leads to the presence of flavor changing neutral currents (FCNCs).

In this work, we consider the flavor violation induced in these models. 
In particular, we study the FCNC effects leading to $b\to s$ transitions
induced by massive color-octet gauge bosons that are present in all the specific 
realizations. In the case of warped extra dimensions, the KK excitations of the gluon
field give rise to  these interactions provided that the zero-mode fermions 
of third generation quarks are localized toward the TeV brane, 
as it is required
in order to obtain a large top Yukawa coupling. In Topcolor theories, the dynamics
responsible for top-condensation is essentially flavor violating. 
In both cases, there might be additional flavor violation through color-singlet
states. For instance, in some Topcolor models there is a complicated
flavor-violating scalar sector, as well as color-singlet gauge bosons. 
In the case of warped extra dimensions, there is the possibility of flavor 
violating KK excitations of the $Z$, plus those of other additional color-singlet
gauge bosons propagating in the bulk, depending on what the ``weak'' gauge symmetry 
is in the bulk. The presence of these extra contributions is 
model-dependent and therefore
we choose to focus on the color-octet gauge boson interactions, which in any event
tend to give the largest effects due to the presence of $\alpha_s$. 
We show that flavor-violating color-octet interactions strongly coupled to the third
generation could naturally lead to large effects in $b\to s\bar ss$ and 
$b\to s\bar dd$ CP asymmetries, such as in $B_d\to\phi K_s$, $B_d\to\eta'K_s$, 
$B_d\to\pi^0 K_s$ among others. The current data from BABAR and BELLE, 
shown in Table~\ref{data},  
signal the possibility of a deviation of the value of $\sin(2\beta)$ as extracted 
in these modes from the one obtained in the $b\to c\bar cs$ golden 
mode $B_d\to J/\psi K_s$.
\begin{table}
\begin{center}
\begin{tabular}{|c|c|c|} \hline\hline
 Mode& $S_f$ & $C_f$ 
\\ \hline
\underline{$B_d\to\phi K_s$} & & \\
BaBar &$0.45\pm0.44$ &$0.38\pm0.39$ \\
BELLE &$-0.96\pm0.51$ &$-0.15\pm0.30$ \\ \hline
\underline{$B_d\to\eta' K_s$} && \\
BaBar &$0.02\pm0.34$ &$-0.10\pm0.22$ \\
BELLE & $0.43\pm0.28$ & $-0.01\pm0.17$ \\ \hline
\underline{$B_d\to K^+K^-K_s$} & & \\
BELLE& $0.51\pm0.26^{+0.18}_{-0.00}$ & $-0.17\pm0.16$ \\\hline
\underline{$B_d\to\pi^0K_s$} & & \\
BaBar & $0.48^{+0.38}_{-0.47}\pm0.11$ & $0.40^{+0.27}_{-0.28}\pm0.10$ \\ \hline
\end{tabular}
\end{center}
\caption{Most recent data on CP asymmetries in penguin dominated modes, as presented in 
Ref.~\cite{s2bwa}. In the SM, we expect 
$S_f\simeq\sin(2\beta)_{\psi Ks}=0.731\pm0.056$, 
and $C_f\simeq0$.} 
\label{data}
\end{table}
Although errors are still large, they are statistics dominated and are expected
to be considerably reduced by the time the experiments accumulate $500~{\rm fb}^{-1}$.

Several authors have considered various possible new physics sources for deviations
in the CP asymmetries of penguin dominated decays~\cite{othernew} as the ones 
in Table~\ref{data}. These are coming from anomalous $Z$ couplings, a strongly coupled
$Z'$ or loop effects involving, for instance, superpartners in some supersymmetric
scenario. The distinct and theoretically motivated 
possibility that the effects are induced by new color-octet
currents has not been considered in the literature. 

In the next section we derive the flavor-violating interactions in 
models of warped extra dimensions. In Section~\ref{tc}, we do the same in 
generic Topcolor models. In Section~\ref{b2s} we predict the CP asymmetry to be
observed in these scenarios in various $B$ decay modes as a function of the model 
parameters. We finally conclude in Section~\ref{conc}.

\section{Warped Extra Dimensions and Flavor Violation}
\label{warped}
Recently Randall and Sundrum proposed the use 
of a non-factorizable geometry in five dimensions~\cite{rs1} as a solution of the 
hierarchy problem.
The metric depends on the five dimensional coordinate $y$ and is given by 
\begin{equation}
ds^2 = e^{-2\sigma(y)} \eta_{\mu\nu} dx^\mu dx^\nu - dy^2~,
\label{metric}
\end{equation} 
where $x^\mu$ are the four dimensional coordinates, $\sigma(y) = k |y|$, with 
$k\sim M_P$ characterizing the curvature scale. The extra dimension is compactified
on an orbifold $S_1/Z_2$ of radius $r$ so that the bulk is a slice of ${\rm AdS}_5$
space between two four-dimensional boundaries. The metric on these boundaries generates
two effective scales: $M_P$ and $M_P e^{-k\pi r}$. In this way, values of 
$r$ not much larger than the Planck length ($kr\simeq (11-12)$)
can be used in order to generate
a scale $\Lambda_r\simeq M_Pe^{-k\pi r}\simeq~{\rm O(TeV)}$ on one of the boundaries.

In the original RS scenario, only gravity was allowed to propagate in the bulk, 
with the Standard Model (SM) fields confined to one of the boundaries. 
The inclusion of matter and gauge fields in the bulk has been extensively treated in the
literature~\cite{bulk1,chang,gn,gp,huber1,dhr1,dhr2}. 
Here we are interested in examining the situation when the SM fields 
are allowed to propagate in the bulk. The exception is the 
Higgs field which must be localized on the TeV boundary in order for 
the $W$ and the $Z$ gauge bosons to get their observed 
masses~\cite{chang}. The gauge content in the bulk may be that of the SM, or it might
be extended to address a variety of model building and phenomenological issues.
For instance, the bulk gauge symmetries may correspond to Grand Unification scenarios, 
or they may be extensions of the SM formulated to restore enough custodial symmetry and 
bring electroweak contributions in line with constraints. 
In addition, and as it was recognized in Ref.\cite{gp}, it is possible to generate
the fermion mass hierarchy from $O(1)$ flavor breaking in the bulk masses of fermions. 
Since bulk fermion masses result in the localization of fermion zero-modes, 
lighter fermions 
should be localized toward the Planck brane, where their wave-function has exponentially
suppressed overlap with the TeV-localized Higgs, whereas fermions with order one
Yukawa couplings should be localized toward the TeV brane. 

This creates an almost inevitable tension: since the lightest KK excitations of gauge
bosons are localized toward the TeV brane, they tend to be strongly coupled to 
zero-mode fermions localized there. Thus, the flavor-breaking fermion localization 
leads to flavor violating interactions of the KK gauge bosons.
In particular, this is the case when one tries to obtain the correct top Yukawa coupling:
the KK excitations of the various gauge bosons propagating in the bulk will 
have FCNC interactions with the third generation quarks, as we will see below in detail.

\begin{figure}
\leavevmode
\centering
\epsfig{file=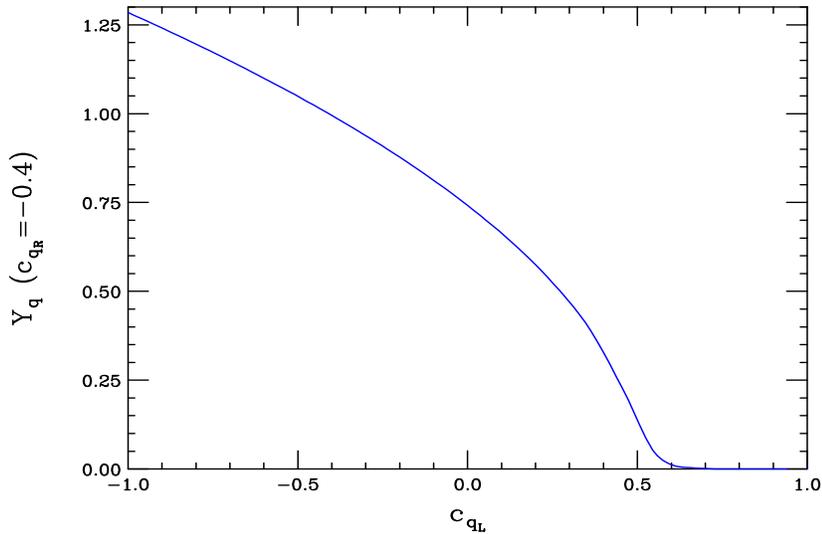,width=7cm,height=10.8cm,angle=90}
\caption{Yukawa coupling of a left-handed zero mode fermion for a TeV-brane localized 
right-haded fermion.
}
\label{yukawa}
\end{figure}

\begin{figure}
\leavevmode
\centering
\epsfig{file=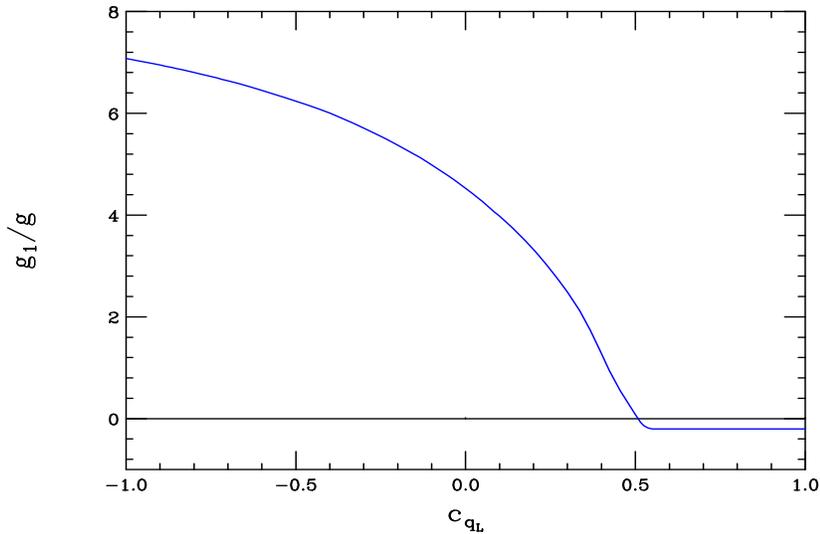,width=7cm,height=10.8cm,angle=90}
\caption{Coupling of the first KK excitation of a gauge boson to a zero mode fermion vs. 
the bulk mass parameter c, normalized to the four-dimensional gauge coupling $g$.  
}
\label{g1g0}
\end{figure}

The action for fermion fields in the bulk is given by~\cite{chang,gn}
\begin{equation}
S_f = \int d^4x~dy~ \sqrt{-g} \left\{ \frac{i}{2}\bar\Psi\hat{\gamma}^M
\left[{\cal D}_M
-\raisebox{0.12in}{$\leftarrow$}\hspace*{-0.18in}{\cal D}_M\right]\Psi   
- {\rm sgn}(y) M_f \bar\Psi\Psi\right\}~,
\label{sfermions} 
\end{equation}
where the covariant derivative in curved space is
\begin{equation}
{\cal D}_M \equiv \partial_M + \frac{1}{8}\,[\gamma^\alpha,~\gamma^\beta]~V_\alpha^N
~V_{\beta N;M}~,
\label{covder}
\end{equation} 
and $\hat{\gamma}^M\equiv V_\alpha^M\,\gamma^\alpha$, with 
$V_\alpha^M={\rm diag}(e^\sigma,e^\sigma,e^\sigma,e^\sigma,1)$ the inverse vierbein.
The bulk mass term $M_f$ in eqn.~(\ref{sfermions}) is expected to be of order
$k\simeq M_P$. 
Although the fermion field $\Psi$ is non-chiral, we can still define 
$\Psi_{L,R}\equiv \frac{1}{2}(1\mp\gamma_5)\Psi$. The KK decomposition can be written as
\begin{equation}
\Psi_{L,R}(x,y) = \frac{1}{\sqrt{2\pi r}}\,\sum_{n=0}\,\psi_n^{L,R}(x) e^{2\sigma} 
f_n^{L,R}(y)~,
\label{kkfermion}
\end{equation}
where $\psi_n^{L,R}(x)$ corresponds to the $nth$ KK fermion excitation and is 
a chiral four-dimensional field. 
The zero mode wave functions are 
\begin{equation}
f_0^{R,L}(y) = \sqrt{\frac{2k\pi r\,(1\pm 2c_{R,L})}{e^{k\pi r(1\pm2c_{R,L})}-1}}\;
e^{\pm c_{R,L} \,k\,y}~,
\label{zeromode} 
\end{equation}
with $c_{R,L}\equiv M_f/k$ parametrizing the bulk fermion mass in units of the 
inverse AdS radius $k$. The $Z_2$ orbifold projection is used so that only 
one of these is actually allowed, either a left-handed or a right-handed zero mode.
The Yukawa couplings of bulk fermions to the TeV brane Higgs can be written as 
\begin{equation}
S_Y = \int d^4x \,dy \,\sqrt{-g}\,\frac{\lambda_{ij}^{5D}}{2\,M_5} \,\bar{\Psi}_i(x,y) 
\delta(y-\pi r) H(x)
\Psi_j(x,y)~,
\label{yuka5d}
\end{equation}
where $\lambda_{ij}^{5D}$ is a dimensionless parameter and $M_5$ is the fundamental 
scale or cutoff
of the theory. Naive dimensional analysis tells us that we should expect 
$\lambda_{ij}^{5D}\simlt 4\pi$. 
Thus the 4D Yukawa couplings as a function of the bulk mass parameters are
\begin{equation}
Y_{ij} = \left(\frac{\lambda_{ij}^{5D}\,k}{M_5}\right)\, 
\sqrt{\frac{(1/2- c_L)}{e^{k\pi r(1-2c_L)}-1}}\,
\sqrt{\frac{(1/2- c_R)}{e^{k\pi r(1-2c_R)}-1}}\,
e^{k\pi r(1-c_L-c_R)}~. 
\label{yuka4d}
\end{equation}
Given that we expect $k\simlt M_5$ then the factor $\lambda_{ij}^{5D}\,k/M_5\simeq O(1)$.
In Figure~\ref{yukawa}, the Yukawa coupling of a left-handed zero-mode fermion is plotted as
a function of $c_L$ for a choice of $c_R=-0.4$, i.e. a TeV-brane localized right-handed zero 
mode.
We see that in order to obtain an $O(1)$ Yukawa coupling, the bulk mass parameter $c_L$ 
should naturally be in the range $[-0.7,0]$. In other words, the left-handed zero-mode
should also be localized toward the TeV brane. This however, posses a problem since
it means that the left-handed doublet $q_L$, and therefore $b_L$ should have a 
rather strong coupling to the first KK excitations   of gauge bosons.
In Figure~\ref{g1g0} we plot the coupling of the first KK excitation of a gauge boson 
to a zero-mode fermion vs. the fermion's bulk mass parameter $c$~\cite{gp}. 
Thus, the localization of the third generation quark doublet $q_L$ in the range $c_L=[-0.7,0]$
leads to potentially large flavor violations, not only with the top quark, but also with 
$b_L$. 

This induced flavor violation of KK gauge bosons with $b_L$ (we assume $b_R$ localized on the 
Planck brane) is, in principle, constrained by the precise measurement of the 
$Z\to b\bar b$ interactions at the $Z$-pole. 
For instance, Ref.\cite{agashe2} considers a 
$SU(3)_c\times~SU(2)_L\times~SU(2)_R\times U(1)_{B-L}$ 
gauge theory in the bulk. After electroweak symmetry breaking the $Z$ mixes with its KK 
excitations, as well as with the KK modes of a $Z'$. This generates $\delta g_L^b\simlt 
O(1\%)g_L^b$, compatible with current bounds  as long as $c_L\simgt 0.3$. 
This still leaves a large flavor violating coupling of the first KK 
excitations to the $b_L$, 
as we can see from Figure~\ref{g1g0}. 
Since in general the weak sector of the bulk gauge theory varies from model to model, we
will focus on the flavor violating effects of the KK gluons. 

In most of these models, only the third generation left-handed quark doublet 
and $t_R$ are required to be localized toward the TeV brane, whereas $b_R$
could in principle be toward the Planck brane together with the light fermions. 
In this case, the KK gluons and other KK gauge excitations would not 
have strong flavor violating couplings with $b_R$. On the other hand, if 
their bulk mass parameter was similar to the one of $t_R$ or $(t~b)^T_L$, 
it would also contribute to flavor violating vertices. 
This, for instance, would be the case if instead of having 
$SU(2)_L\times~SU(2)_R$ in the bulk, one has $SO(4)$ as the ``weak''
gauge symmetry~\cite{wcsaki}. In this case, $(t~b)^T_L$ and $(t~b)^T_R$
should have the same bulk mass.
Although the case with an exact $SO(4)$ symmetry may not be phenomenologically viable due 
to constraints from electroweak precision 
observables~\cite{ynnh}, an approximate $SO(4)$
could be accommodated. This would still lead to a $b_R$ zero-mode with a TeV-localized 
wave function, and to interesting phenomenology, both in $B$ physics as well 
as in electroweak observables.

\section{Topcolor Models and Flavor Violation}
\label{tc}
In Topcolor theories, the flavor violating interactions are intended to 
become supercritical leading to  $\langle\bar t_Lt_R\rangle\not=0$, therefore
providing a backdrop for top-condensation in a broken gauge theory.
The minimal module of a Topcolor model can be viewed as the breaking
of $SU(3)_1\times SU(3)_2\longrightarrow SU(3)_c$ at a $O({\rm TeV})$ scale.
This leaves not only the usual massless gluons but also a color-octet massive 
gluon with a typical $O({\rm TeV})$ mass.
Third generation left handed quark doublet $(t~b)^T_L$ and $t_R$ must transform under
the stronger $SU(3)_1$, whereas the lighter generations transform under the weaker 
$SU(2)_2$. The $b_R$ may transform under the stronger group or not depending on the 
choice of ``weak'' gauge group (Notice the similarity with the 
situation in the previous section). Fermions that transform under $SU(3)_1$ 
will couple to the heavy gluons with an enhanced coupling given by 
$\simeq (g_1/g_2) g_c$, with $g_c$ the 
standard color coupling. The lighter fermions, on the other hand, couple like
$(g_2/g_1) g_c$. 
These non-universal couplings lead to tree-level FCNCs which are particularly important
in observables involving the third generation~\cite{bbhk}. In particular, 
the exchange of the heavy gluon would lead to four-fermion operators mediating
transitions such as $b\to s\bar qq$ with strength $\alpha_s$. Thus, they will be comparable
to gluonic penguins resulting in potentially important 
effects in the CP asymmetries of penguin dominated modes, as we 
see in the next section.

\section{Signals in $b\to s$ Hadronic Decays} 
\label{b2s}

As mentioned above, the flavor-changing exchange of KK gluons leads 
to four-fermion 
interactions contributing to the quark level processes $b\to d\bar qq$ and 
$b\to s\bar qq$,
with  $q=u,d,s$. 
We are interested in contributions that are typically of $\simeq~\alpha_s$ strength
due to the fact that in the product of a third generation current times a lighter 
quark current the enhancement in the former is (at least partially) canceled by 
the suppression of the latter.
At low energies, the $b\to d_i\bar qq$ processes are  described by the 
effective Hamiltonian~\cite{bbl}
\begin{equation}
{\cal H}{\rm eff.} = \frac{4G_F}{\sqrt{2}}\; V_{ub}V_{ui}^*\left[C_1(\mu) O_1
+C_2(\mu) O_2\right]
-\frac{4G_F}{\sqrt{2}}\;V_{tb}V_{ti}^*\;\sum_{j=3}^{10}\, C_i(\mu) O_i
+{\rm h.c.},
\label{heff}
\end{equation}
where $i=d,s$ and the operator basis can be found in Ref.~\cite{bbl}.

The Wilson coefficients $\{C_j(\mu)\}$ contain the short-distance information
which, in the SM, arises from integrating out heavy particles such as the
$W$ and $Z$ gauge bosons and the top quark. 
In the SM, the operators $\{O_3-O_6\}$ are generated from one-loop
gluonic penguin diagrams, whereas operators $\{O_7-O_{10}\}$ arise from 
one-loop electroweak penguin diagrams. 
The Hamiltonian describing the $b\to s\bar qq$ decays is obtained by 
replacing $V_{ts}^*$ for $V_{td}^*$ in eq.~(\ref{heff}). 
Contributions from physics beyond the SM affect the Wilson coefficients 
at some high energy scale.
Additionally, new physics could generate low energy interactions
with the ``wrong chirality'' with respect to the SM. This would  expand the 
operator basis to include operators of the form $(\bar s_R \Gamma b_R) 
(\bar q_{\lambda} \Gamma q_{\lambda})$, where $\Gamma$ reflects the 
Dirac and color 
structure and $\lambda=L,R$. 

The exchange of color-octet gauge bosons
such as KK gluons of the Randall-Sundrum scenario described in Section~\ref{warped} 
or  the top-gluons of Section~\ref{tc} 
generate flavor-violating currents with  the third generation quarks. 
Upon diagonalization of the Yukawa matrix, this results in FCNCs at tree level
due to the absence of a complete GIM cancellation. The off-diagonal 
elements of the left and right, up and down quark rotation matrices $U_{L,R}$
and $D_{L,R}$ determine the strength of the flavor violation.
In the standard model only the left-handed rotations are observable
through $V_{\rm CKM}=U^\dagger_LD_L$. Here,  $D_{L,R}^{bs}$, 
$D_{L,R}^{bd}$,  $U_{L,R}^{tc}$, etc., become actual observables.

The tree-level flavor changing interactions induced by the color-octet exchange are 
described by a new addition to the effective Hamiltonian that in 
general can be written, for $b\to s $ transitions, as
\begin{equation}
\delta{\cal H}_{\rm eff.} = \frac{4\pi\alpha_s}{M_G^2}\,
D_L^{bb*}\,D_L^{bs}\,|D_L^{qq}|^2\,e^{-i\omega}
\;\chi\;(\bar s_L\gamma_\mu T^a b_L)\;(\bar q_L\gamma^\mu
 T^a q_L)+{\rm h.c.}~.
\label{dheff}
\end{equation}
where $\omega$ is the phase relative to  
he SM contribution; and 
$\chi\simeq O(1)$ is a model-dependent parameter. 
For instance in the Randall-Sundrum case, $\chi=1$ corresponds to the 
choice of $c_L^b\simeq 0$ that gives a coupling of the KK gauge boson 
about five times larger than the corresponding standard model value for that
gauge coupling. 
On the other hand, for Topcolor models $\chi=1$, as noted in Section~\ref{tc}. 
Eq.~(\ref{dheff}) shows the 
case of left-handed flavor violation interactions. However, in general right-handed 
flavor violating vertices, leading to ``wrong-chirality'' operators,  
can also be present. This is the case, for instance in 
the Randall-Sundrum scenario where the right-handed b quark is also 
localized toward the TeV brane. It is also the case in most Topcolor
models. 
An expression analogous to (\ref{dheff}) is obtained by replacing 
$d$ for $s$ in it. This would induce effects in $b\to d$ processes.

From eq.~(\ref{dheff}) 
we can see that 
the color-octet exchange 
generates contributions to all gluonic penguin operators. 
Assuming that the diagonal factors obey $|D_L^{qq}|\simeq 1$, 
these will have the form
\begin{equation}
\delta C_i=-\pi\alpha_s(M_G)\;\left(\frac{v}{M_G}\right)^2\;\left|\frac{D_L^{bs*}}
{V_{tb}V_{ts}^*}\right|\,e^{-i\omega}\;f_i~\chi~,
\label{dci}
\end{equation}
where  $f_3=f_5=-1/3$ and $f_4=f_6=1$, and $v=246~$GeV. 
This represents a shift in the Wilson coefficients at the high scale. We then must evolve
the new coefficients down to $\mu=m_b$ by making use of the renormalization group
evolution~\cite{bbl}. 
The effects described by eq~(\ref{dci}) are somewhat
diluted in the final answer due to a large contribution from the mixing with $O_2$. 
Still, potentially large effects remain. 

The phase $\omega$ in eq.~(\ref{dheff}) is in principle a free parameter in most
models and it could be large. This is even true in the left-handed sector, 
since in general $V_{\rm CKM}$ comes from both the up and down quark rotation. 
Only if we were to argue that all of the CKM matrix comes from the down sector
we could guarantee that $\omega=0$. Furthermore, there is no such constraint in the
right-handed quark sector.

We now examine what kind of effects these flavor violating
terms could produce. Their typical strength is given by $\alpha_s$ times
a CKM-like factor coming from the $D_{L,R}$ off-diagonal elements connecting to $b$. 
The fact that the coupling is  tree-level is somewhat compensated 
by the suppression factor $(v/M_G)^2$, for $M_G\simeq O(1)~$TeV. 
Still, the contributions of
eq.~(\ref{dheff}) are typically larger than the SM Wilson coefficients at the 
scale $M_W$, and in fact are comparable
to the Wilson coefficients at the scale $m_b$.  
Therefore, they could significantly affect
both the rates and the CP asymmetries.
In what follows we will make use of $\delta C_i(m_b)\simeq \delta C_i(M_W)$, 
neglecting the renormalization group effects in these four-quark operators.
These typically will result in logarithmic enhancements in terms of 
$\ln(M_G^2/\mu^2)$, with $\mu$ some low energy scale. These effects should be resummed
by the renormalization group. We ignore them here in order to make an estimate of the 
deviations caused by (\ref{dci}).

We are interested in examining the effects of eq.~(\ref{dheff}) in 
$b\to s\bar ss$ and $b\to s\bar dd$ pure penguin processes, such as 
$B_d\to\phi K_s$, $B_d\to\eta' K_s$ and $B_d\to \pi^0 K_s$ in light of the fact
that these modes contain only small tree-level contamination of standard model
amplitudes. All these decays, constitute a potentially clean test of the standard
model since their CP asymmetries are predicted to be a measurement of 
$\sin(2\beta)_{\psi K_s}$, the same
angle of the unitarity triangle as in the $b\to c\bar cs$ 
tree-level processes such as $B_d\to \psi K_s$, up to small corrections~\cite{corr}.

In order to estimate these effects and compare them to the current experimental 
information on these decay modes we will compute the matrix elements
of ${\cal H}_{\rm eff.}$ in the factorization 
approximation~\cite{fac} as described in Reference~\cite{ali}. 
Although the predictions for the branching ratios suffer from significant uncertainties,
we expect that these largely cancel 
when considering the effects in the CP asymmetries. (We examine this expectation
below in detail).
Thus the CP asymmetries
in non-leptonic $b\to s$ penguin dominated processes constitute a suitable
set of observables to test the effects of these color-octet states. 
As we will see, these observables are the natural place for a first observation 
of this physics.  

\begin{figure}[t]
\leavevmode
\centering
\epsfig{file=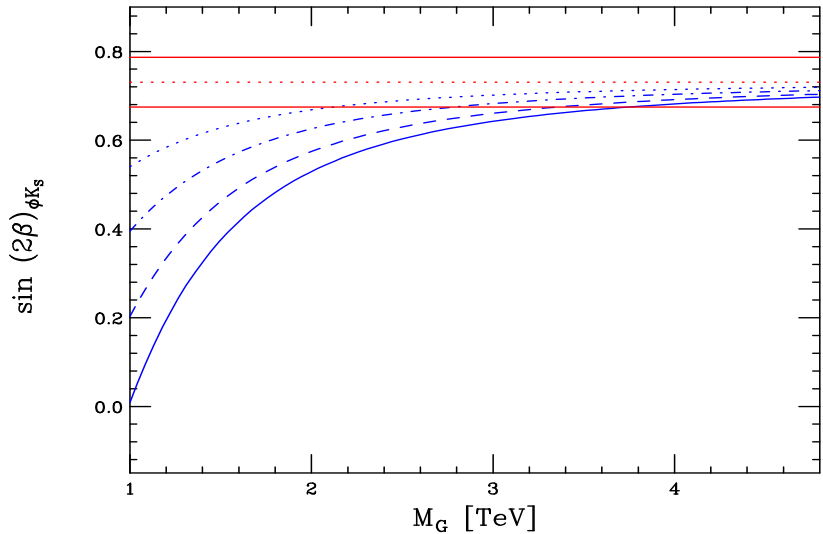,width=7cm,height=10.8cm,angle=90}
\caption{
The quantity to be extracted from the CP violation asymmetry in 
$B_d^0\to\phi K_s$ vs. the heavy gluon mass and for various values 
of the decay amplitude phase $\omega$. The curves correspond
to $\pi/3$ (solid), $\pi/4$ (dashed) and $\pi/6$ 
(dot-dash), and $\pi/10$ (dotted). 
The horizontal band corresponds to the world average value~\cite{s2bwa}
as extracted from $B_d\to J/\psi K_s$, 
$sin(2\beta)_{\psi K_s}=0.731\pm 0.056$. 
}
\label{sin2bt}
\end{figure}

In Fig.~\ref{sin2bt} we plot $\sin(2\beta)_{\phi K_s}$ vs. 
the KK gluon mass and for various values of the phase $\omega$.
Here, for concreteness, we have taken $|D_L^{bs}|=|V^*_{tb}V_{ts}|$, 
assumed $b_R$ is localized on the Planck brane, 
and $\chi=1$ in order to illustrate the size of the effect. 
The horizontal band corresponds to the $B_d\to\psi K_s$ measurement, 
$\sin(2\beta)_{\psi K_s}=0.731\pm0.056$~\cite{s2bwa}. 
Shown are only positive values of $\omega$, as negative ones increase the 
value of $\sin(2\beta)$ contrary to the trend in the data. 
We see that sizeable deviations from the SM expectation are present 
for values in the region of interest  $M_G\simgt 1$~TeV.
This will be the case as long as $|D_L^{bs}|\simeq |V_{ts}|$, 
and $\chi\simeq O(1)$, both natural assumptions.  
For $D_L^{bs}$, this is valid as long as a significant fraction 
of the corresponding CKM elements comes from the down-quark rotation. 
On the other hand, $\chi\simeq O(1)$ in all the models considered here. 
In addition, we have not considered the effects of of $D_R^{bs}$, which 
could make the effects even larger. 

\begin{figure}[t]
\leavevmode
\centering

\epsfig{file=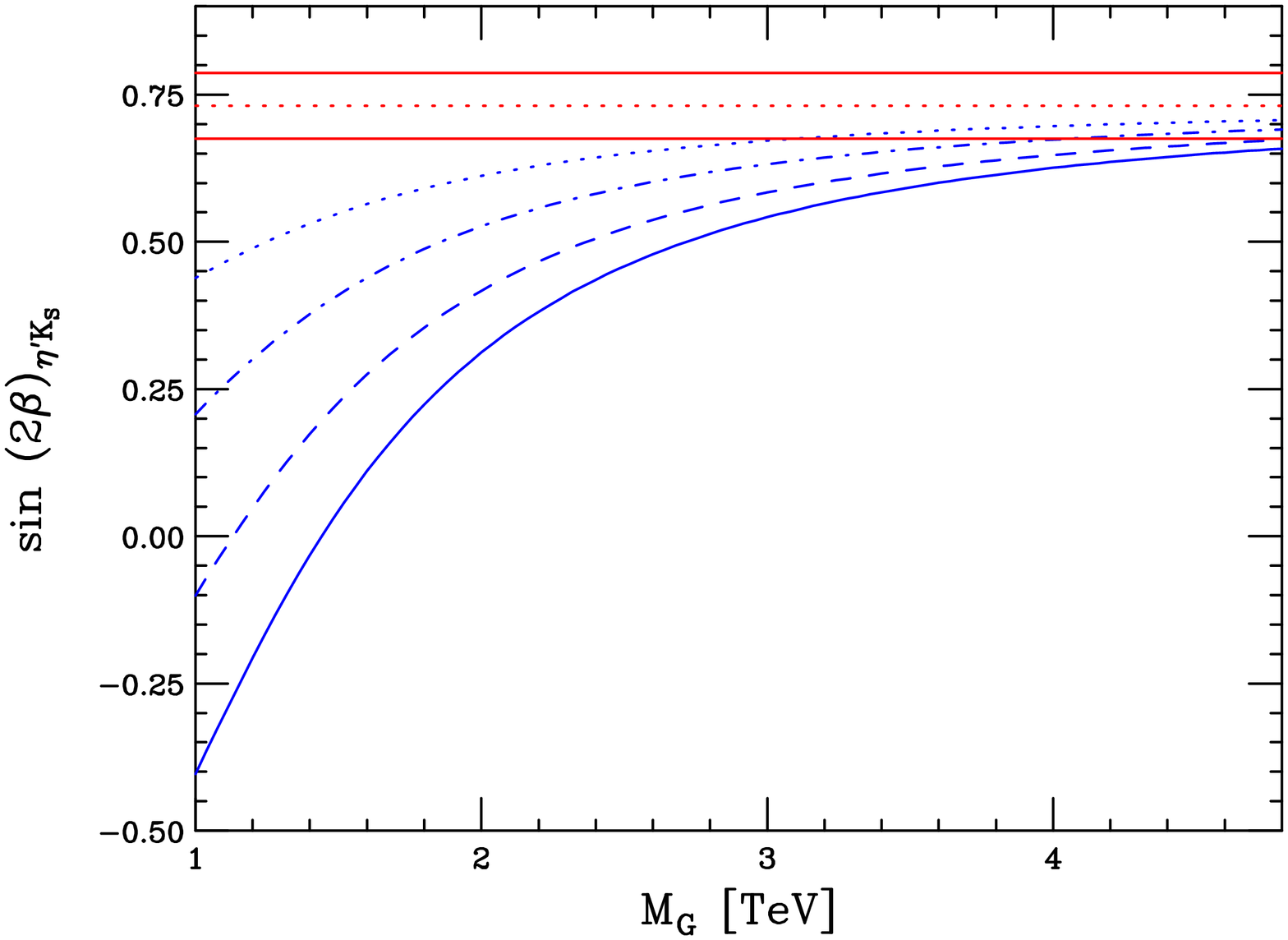,width=7cm,height=10.8cm,angle=90}
\caption{
The quantity to be extracted from the CP violation asymmetry in 
$B_d^0\to\eta'K_s$ vs. the heavy gluon mass and for various values 
of the decay amplitude phase $\omega$. The curves correspond
to $\pi/3$ (solid), $\pi/4$ (dashed) and $\pi/6$ 
(dot-dash), and $\pi/10$ (dotted). 
The horizontal band corresponds to the world average value~\cite{s2bwa}
as extracted from $B_d\to J/\psi K_s$, 
$sin(2\beta)_{\psi K_s}=0.731\pm 0.056$. 
}
\label{etap_s2b}
\end{figure}

Similarly, the decay $B_d\to\eta'K_s$ is also dominated by the 
$b\to s\bar ss$ penguin contribution. 
In Fig.~\ref{etap_s2b} we plot $\sin(2\beta)_{\eta'K_s}$ vs. 
the KK gluon mass and for various values of the phase $\omega$, for the same
choice of parameters as for the previous case. 
Although the effect in this mode appears even bigger, there are several 
additional sources of uncertainties. First, within the factorization approximation
the effect in $\sin(2\beta)_{\eta' K_s}$ depends on the quantity
\begin{equation}
r\equiv\left(\frac{f^u_{\eta'}}{f_K}\right)\;\left(\frac{f^{B\to K}(m^2_{\eta'})}
{f^{B\to \eta'}(m^2_{K})}\right)~,
\label{rdef}
\end{equation}
where $if^q_{\eta'}\,p_\mu=\langle 0|\bar q\gamma_\mu\gamma_5 q|\eta'\rangle$ is 
the $\eta'$ decay constant through the $q=u,d,s$ axial-vector current, and 
$f^{B\to P}(q^2)$ are the semileptonic form-factors for the $B\to P$ decays.
In Fig.~\ref{etap_s2b} we take the values for the corresponding parameters 
from Ref.~\cite{ali}, which results in $r\simeq 1$. 
However, the answer is very sensitive to small variations in $r$, which, for instance, 
could be provided by  our 
limited understanding of the $B\to\eta'$ form-factor in eq.~(\ref{rdef}).
In addition to this source of uncertainty, it is generally expected that
this decay mode would receive an enhancement from the large coupling of the 
QCD anomaly~\cite{anomaly} to the singlet piece in $\eta'$. Several attempts have been
made to estimate the size of this effect, suggested first by the large branching 
fractions with $\eta'$ in the final states. The anomaly contribution not only 
could introduce an uncertainty of order one in the calculation of the 
effects in the CP asymmetry, but it might also dilute new physics contributions
as long as these do not arise from a modification of the $b\to sg$ vertex.

\begin{figure}[t]
\leavevmode
\centering

\epsfig{file=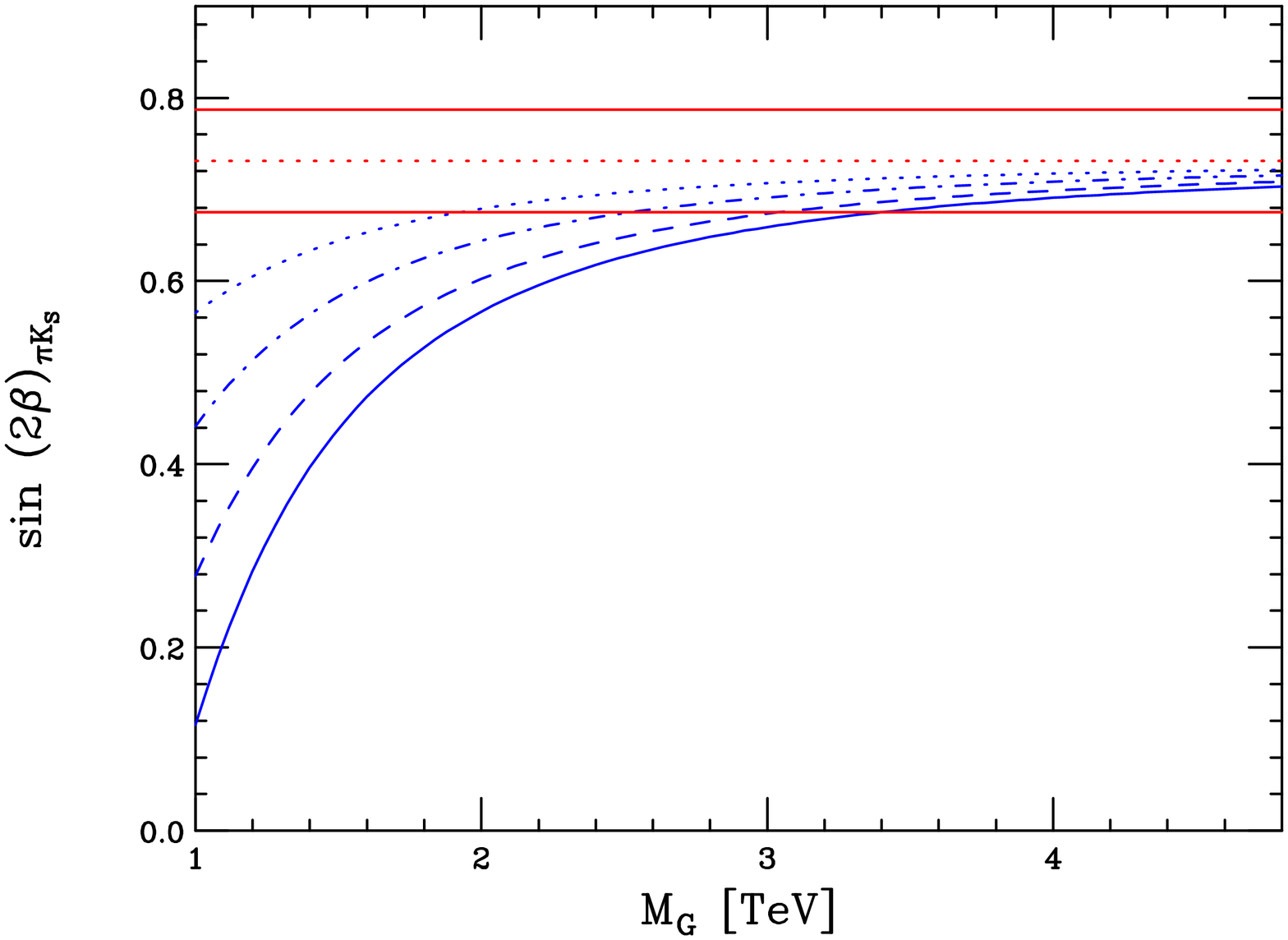,width=7cm,height=10.8cm,angle=90}
\caption{
The quantity to be extracted from the CP violation asymmetry in 
$B_d^0\to\pi^0K_s$ vs. the heavy gluon mass and for various values 
of the decay amplitude phase $\omega$. The curves correspond
to $\pi/3$ (solid), $\pi/4$ (dashed) and $\pi/6$ 
(dot-dash), and $\pi/10$ (dotted). 
The horizontal band corresponds to the world average value~\cite{s2bwa}
as extracted from $B_d\to J/\psi K_s$, 
$sin(2\beta)_{\psi K_s}=0.731\pm 0.056$. 
}
\label{piks_s2b}
\end{figure}

We finally consider the effects of the KK gluons in the $b\to s\bar dd$ mode
$B_d\to\pi^0 K_s$. Here, just as for $B_d\to\phi K_s$, the CP asymmetry can be 
rather cleanly predicted in the factorization approximation and uncertainties are
considerably smaller. In Fig.~\ref{piks_s2b} we plot $\sin(2\beta)_{\pi^0 K_s}$ 
vs. he KK gluon mass for various choices of $\omega$. 
The effect in general appears to be somewhat smaller than the ones in 
$B_d\to\phi K_s$ and $B_d\to\eta' K_s$ for the same choice of $M_G$ and $\omega$. 

The flavor violating  exchange of the KK gluon also induces
an extremely large contribution to $B_s-\bar{B}_s$ mixing. 
This is roughly given by 
\begin{equation}
\Delta m_{B_s} \simeq 200 {\rm ps}^{-1}\,\left(\frac{|D_L^{bs}|}{\lambda^2}\right)^2\;
\left(\frac{2~{\rm TeV}}{M_G}\right)^2\;\left(\frac{g_{10}}{5}\right)^2~,
\label{dmbs}
\end{equation}
where $\lambda\simeq 0.22$ is the Cabibbo angle, and  
$g_{10}\equiv g_1/g$ represents the enhancement of the zero-mode fermion 
coupling to the first KK gluon with respect to the four-dimensional gauge coupling, 
as plotted in Fig.~\ref{g1g0}. 
The contribution of eq.~(\ref{dmbs}) by itself is  about $10$ times larger than 
the SM one for this natural choice 
of parameters, and would deem $B_s$ oscillations too rapid for observation 
at the Tevatron and other similar experiments. 

There are also similar contributions to $\Delta m_{B_d}$, when $D_L^{bs}$ is replaced 
by $D_L^{bd}$. These were examined in Ref.~\cite{blr} in the context of Topcolor
assisted technicolor, a much more constrained brand of Topcolor than the one we 
consider here. The bounds found in Ref.~\cite{blr} can be accommodated as long 
as $|D_L^{bd}|\simlt |V_{td}|$ which is not a very strong constraint.

Thus, we see that the flavor-violation 
effects of the first KK gluon excitation in Randal-Sundrum 
scenarios where the $SU(3)_c$ fields propagate in the bulk can be significant in 
non-leptonic $B$ decays and specifically in their CP asymmetries. The dominance of these
effects over those induced by ``weak'' KK excitations, such as KK $Z$ and
$Z'$'s, due to the larger coupling would explain the absence of any effects
in $b\to s \ell^+\ell^-$ processes, where up to now the data is consistent with 
SM expectations~\cite{bsll}. Deviations in the CP asymmetries of $b\to s$ non-leptonic
processes would naturally be the first signal of new physics in these scenarios. 
These very same effects can be obtained by the exchange of the heavy gluons
present in generic Topcolor models.

\section{Conclusions}
\label{conc}
We have shown that in Randall-Sundrum scenarios with gauge and matter fields in the 
bulk, naturally occurring flavor violation is likely to be first observed in 
non-leptonic $B$ decays. Specifically, in models where obtaining a 
suitable top Yukawa coupling requires the localization of the third generation 
left handed doublet and right handed top quark  toward the TeV brane,
the non-universal interactions of the KK gluons result in tree-level 
FCNCs inducing $b\to s q\bar q$ decays with ($q=d,s$). 
In addition, the localization of $b_R$ toward the TeV brane results in extra
contributions from flavor changing right handed currents.

We have studied the effects in the CP asymmetries in $B_d\to\phi K_s$, $B_d\to\eta' K_s$
and $B_d\to\pi^0 K_S$, where data is already available. In these three modes 
it is expected that the CP asymmetry measurements yield the same value 
of $\sin 2\beta$ as in 
the vastly studied tree-level decay $B_d\to J/\psi K_s$. We have seen that for naturally
occurring order one relative phases, and for KK gluon masses in the few TeV range, large
deviations in the CP asymmetries are expected, as it can be seen in 
Figs.~\ref{sin2bt},~\ref{etap_s2b}~and~\ref{piks_s2b}.  
Similar deviations should also be expected in $B_d\to K^+K^-K_s$, as 
well as in a host of $B_s$ decay modes, where the asymmetry is expected
to be negligibly small in the SM.  
New data to be available in the next few years from BaBar and BELLE might 
strengthen the case for new physics in these modes beyond the 
data in Table~\ref{data}. This, coupled with increasingly  precise observations in 
$b\to s\ell^+\ell^-$ and $b\to s\gamma$ in agreement with SM predictions, 
would point in the direction 
of a massive color-octet state strongly coupled to the third generation quarks as
the source of the deviation. 

We have seen that these effects can also be obtained in generic Topcolor
models. It is not possible to distinguish these two sources from $B$ physics
alone. This is true of any color-octet flavor-violating gauge interaction 
that couples strongly to the third generation. Here we focused on these two interesting
cases, KK gluons in Randall-Sundrum scenarios and Topcolor. Other model building 
avenues addressing fermion masses might result in similar effects.
In addition, the large contributions to $B_s$ mixing, perhaps rendering $\Delta m_{B_s}$ 
{\em too large } to be observed, is an inescapable prediction in this scenario, as 
it can be seen in eq~(\ref{dmbs})\footnote{This feature is not unique of this scenario and 
it is likely to be present in many new physics scenarios giving large effects in 
$b\to s$ non-leptonic decays.}.

Finally, there will also be contributions from the heavy gluons to 
other non-leptonic $B$ decays, such as $B\to\pi\pi$, etc. These modes have 
less clean SM predictions. However, if the deviations hinted in Table~\ref{data}
are confirmed by data samples of $500{\rm fb}^{-1}$, to be accumulated in the 
next few years, it might prove of great importance to confirm the existence 
of these effects in less clean modes, perhaps requiring even larger data samples.
Even if the heavy gluons are directly observed at the LHC, their flavor-violating
interactions will be less obvious there than from large enough $B$ physics samples.
Thus, if flavor violating interactions are observed at the LHC, high precision 
$B$ physics experiments could prove crucial to elucidate what is their role 
in fermion mass generation.

\vskip1.0cm
\noindent
{\bf Acknowledgments}
The author thanks Z.~Chacko, G.~Hiller and J.~Terning for useful discussions. 
He is also grateful to the Aspen Center for Physics for its hospitality during part 
of this work.
This work was supported by the Director, Office of Science, 
Office of High Energy and Nuclear Physics of the U.S. Department of 
Energy under Contract DE-AC0376SF00098.

\end{document}